\documentclass[aps,prb,twocolumn,floatfix,showpacs,amsmath,amssymb]{revtex4}
\usepackage{graphicx}
\usepackage[hypertex]{hyperref}

\begin{document}

\newlength\smallfigwidth
\smallfigwidth=3.2 in

\preprint{KSU/UFV Nov. 2004}

\title{Vacancy effects in an easy-plane Heisenberg model:
reduction of $T_c$ and doubly-charged vortices}
\author{G.\ M.\  Wysin}
\email{wysin@phys.ksu.edu}
\homepage{http://www.phys.ksu.edu/~wysin}
\affiliation{
Departamento de F\'isica,
Universidade Federal de Vi\c cosa, 
Vi\c cosa, 36570-000, Minas Gerais, Brazil
}
\altaffiliation{
Permanent address:
Department of Physics,
Kansas State University,
Manhattan, KS 66506-2601
}
\date{November 8, 2004}

\begin{abstract} 
Magnetic vortices in thermal equilibrium in two-dimensional magnets are 
studied here under the presence of a low concentration of nonmagnetic
impurities (spin vacancies).
A nearest-neighbor Heisenberg (XXZ) spin model with easy-plane exchange 
anisotropy is used to determine static thermodynamic properties and vortex 
densities via cluster/over-relaxation Monte Carlo.
Especially at low temperature, a large fraction of the thermally 
generated vortices nucleate centered on vacancies, where they have
a lower energy of formation.
These facts are responsible for the reduction of the vortex-unbinding 
transition temperature with increasing vacancy concentration, similar to
that seen in the planar rotator model.
Spin vacancies also present the possibility of a new effect, namely, the 
appearance of vortices with double topological charges ($\pm 4\pi$ 
change in in-plane spin angle), stable only when centered on vacancies.  
\end{abstract} 

\maketitle
\section{Introduction}
The vortex-unbinding transition in two-dimensional (2D) spin models 
with planar symmetry (Berezinskii-Kosterlitz-Thouless transition
\cite{Berezinskii70,Berezinskii72,Kosterlitz73})
has attracted interest recently with respect to the influence of 
nonmagnetic impurities or spin vacancies in the lattice.
In any real physical system, some fraction of the atoms could be
substituted by impurities, and if these are nonmagnetic, the spins
neighboring the impurity will be strongly affected by the missing
exchange interactions.
Not only could missing bonds cause locally lower energy densities,
but they give the neighboring spins more freedom of motion, which
can be expected to increase the local spin fluctuations.
This can be expected to affect the static configurations, the
thermal equilibrium properties, and the dynamic correlations,
such as in EPR measurements on 
antiferromagnets.\cite{Subbaraman+98,Zaspel+02}

Significant vacancy effects on the static vortex (or antivortex) configurations
of ferromagnets (and antiferromagnets with two sublattices) have already been 
found for a 2D easy-plane Heisenberg model (three spin components).
Zaspel \textit{et al.}\cite{Zaspel+96} found that the critical anisotropy 
strength ($\delta_c\equiv 1-\lambda_c$, see Hamiltonian below) needed to 
stabilize a vortex in the planar configuration on a square lattice is reduced 
from $\delta_c\approx 0.2966$  to the much lower value 
$\delta_{\textrm{cv}}\approx 0.0429$ when the vortex is centered on 
a vacancy.
Wysin\cite{Wysin03} found a similar result at higher precision
($\delta_{\textrm{cv}}\approx 0.0455$), and determined that a 
vacancy at the center of a circular system with free boundaries
produces an attractive potential for a vortex.
Using dynamic relaxation and Monte Carlo simulations, Pereira 
\textit{et al.}\cite{Pereira+03} found that a single vacancy in
a square system with antiperiodic bounary conditions
provides an attractive potential for a vortex.
These works demonstrated a significant energy reduction for 
a vortex formed on a vacancy, compared to one formed in the center of 
a cell of the lattice, whose value depends on the type of lattice
and the boundary conditions.
The resulting vortex-on-vacancy binding energy was found to increase 
with increasing easy-plane anisotropy strength.
Both analytic and numerical calculations by Paula 
\textit{et al.}\cite{Paula04} show that holes cut out of a spin lattice 
similarly produce interesting attractive effects on vortices.

Continuum model calculations for the closely related planar rotator 
model\cite{Mol+02,Leonel+03}  were interpreted to suggest a repulsive 
potential between a planar vortex and a nonmagnetic impurity, however,
this seems contradictory to later calculations.
Studies of a 2D isotropic Heisenberg antiferromagnet 
by M\'ol \textit{et al.}\cite{Mol+03} and Pereira and Pires\cite{Pereira03} 
found oscillatory dynamic modes of solitons pinned to vacancies,  confirming
the presence of an attractive restoring potential.
Considering these most recent calculations for several 
models,\cite{Wysin03,Pereira+03,Paula04,Mol+03,Pereira03} in general
it has been seen that a spin vacancy attracts vortices (or antivortices) and 
lowers their energy of formation.

In terms of the equilibrium thermodynamics, the effect of a concentration of
vacancies on the BKT transition temperature $T_c$ of the easy-plane Heisenberg 
model has not been studied.
On the other hand, Leonel \textit{et al.}\cite{Leonel+03} performed Monte
Carlo (MC) simulations of the planar rotator model (two-component
spins) and found a lowering of the transition temperature with increasing
vacancy density.
It was argued that vacancies produce an effective repulsive potential 
for vortices, thereby increasing the nucleation of pairs and
lowering the transition temperature, but the vortex density was
not measured in the MC simulations.
Using the helicity modulus to determine $T_c$, they found that $T_c$ goes
to zero when the vacancy concentration of a square lattice reaches about 30\%.
A similar lowering of $T_c$ also appears in the MC simulations of Berche 
\textit{et al.}\cite{Berche+03} for the same model, determined by fitting 
the exponent of the spin-spin correlation function to the critical point
value, $\eta=1/4$.
These latter authors found that $T_c$ did not fall to zero until the vacancy 
concentration reached 41\%, a number related to the percolation threshold
for a square lattice.  
In a related bond-diluted planar rotator model, Castro 
\textit{et al.}\cite{Castro+02} used a self-consistent harmonic approximation
with vortex corrections, determining the reduction of $T_c$ with dilution,
and the temperature variation of the correlation function and its 
exponent $\eta$. 

Here it is interesting to consider whether a similar vacancy-induced 
reduction of $T_c$ occurs in the anisotropic Heisenberg model, which is 
a more realistic spin model that has a true time dynamics.
An analysis of the vortex densities in thermal equilibrium, in the presence 
of vacancies,  helps to explain the role of vacancies in generating 
spin disorder around the transition temperature.
The vortices in the anisotropic Heisenberg model also can be expected to 
have planar or out-of-plane structure, depending on the anisotropy strength.
\cite{Hikami80,Takeno80a,Takeno80b,Wysin88,Gouvea+89,Wysin94,Wysin98}
At stronger anisotropy [i.e., for the XY model, $\lambda=0$, see Eq.\
(\ref{Ham})], the stable static vortices are planar, whether pinned on 
vacancies or free from the vacancies.\cite{Zaspel+96,Wysin03}  
Alterntively, at weak anisotropy, both the stable pinned and free vortices
have nonzero out-of-plane spin components, which might be expected to 
significantly modify some equilibrium properties as well as dynamic
correlations.
Therefore, here we present MC simulations for three different anisotropies,
calculating the changes in $T_c$ and the behavior of the vortex densities,
as well as other thermodynamic properties.

It has been customary only to search for singly charged vortices appearing
in MC simulations of pure easy-plane spin systems. 
Looking in individual unit cells (plaquettes) of the lattice, a net
rotation of the in-plane spin angles through $\pm 2\pi$ as one moves around
the cell indicates the presence of a singly-charged vortex ($q=\pm 1$).
When vacancies are present, however, the searching for vorticity must
be modified. 
Here, we searched for net vorticity also in the four unit cells surrounding
any vacancy of the square lattice.  
This allows for the appearance of a new effect, namely, the presence of
$q=\pm2$ vortices,  which always form centered on the vacancies. 
They appear as a very small fraction of the total vorticity density,
and are present regardless of the anisotropy strength.  
Apparently, by pinning on vacancies,  $q=\pm 2$ vorticies lower their energy 
sufficiently due to the missing spin site, leading to greater ease in their
thermal formation.
In addition, at low temperatures, it is found that most vortices (either
$q=\pm 1$ or $q=\pm 2$) form initially on the vacancies, which gives an 
interesting view of how vacancies modify and even control the BKT transition.

After further definition of the model, we describe the MC simulations, 
determinations of $T_c$ using finite-size scaling of the in-plane susceptibility, 
and the vacancy effects at various anisotropies. 
This is followed by some preliminary analysis of the stability properties of 
the doubly-charged vorticies.

\section{Easy-plane model with random repulsive vacancies}
The model to be investigated has classical three-component spins defined at 
the sites $\mathbf{n}$ of a 2D square lattice with unit lattice constant.
The spins can be analyzed either in terms of their Cartesian components
or using polar spherical coordinate angles,
$\vec{S}=(S^x,S^y,S^z) = 
S(\sin\theta \cos\phi, \sin\theta \sin\phi, \cos\theta)$.
The system is an $L\times L$ square with periodic boundary conditions.
We considered $L$ ranging from 16 to 128, using the dependence of
the thermodynamic averages on $L$ to get estimates of the
critical temperature in the infinite size system.

A small vacancy density $\rho_{\textrm{vac}}$ is introduced into the
lattice as follows.
An occupation number $p_{\mathbf n}$ for each site is set to the static 
values $1$ or $0$ depending on whether the site $\mathbf{n}$ is occupied 
by a spin or is vacant.
The fraction $\rho_{\textrm{vac}}$ of the sites has $p_{\mathbf n}$ set
to zero.
(Equivalently, one can keep the spins at the vacant sites but set their
lengths to zero.)
In order to have the most simplified situation,  the vacant sites are
chosen randomly, but no two are allowed to be within the second nearest 
neighbor distance of $\sqrt{2}$ (the diagonal separation across a unit 
cell of the lattice).
In this way, the immediate neighborhoods of all vacancies are equivalent:
each vacant site is surrounded by eight occupied sites.  
This condition greatly simplifies the algorithm for searching for
localized vorticity around the vacant sites.
On the other hand, it limits the possible density of vacancies to be less 
than 0.25 of the lattice sites (achieved in the ordered configuration
having alternating rows of the lattice fully and half occupied by spins).  
In actual practice, by choosing the vacant sites randomly and enforcing 
this constraint (i.e., quenched repulsive vacancies), the maximum achievable 
vacancy density is $\rho_{\textrm{vac}}\approx 0.1872$ .
As a result, a vacancy density needed to push the BKT transition temperature 
down to zero cannot be achieved, and we do not consider this aspect of the
model here.
Instead, we are more interested in the role the vacancies play in 
controlling where the vortices are forming.

Nearest neighbor unit length spins ($S=1$) in this model interact 
ferromagnetically (exchange constant $J>0$) according to a Hamiltonian 
with easy-plane anisotropy specified by parameter $\lambda$,
\begin{equation} \label{Ham}
H = \frac{-J}{2} \sum_{\mathbf{n, a}} 
p_{\mathbf{n}} p_{\mathbf{n+a}} 
\Bigl[ 
S^x_{\mathbf{n}} S^x_{\mathbf{n+a}}+S^y_{\mathbf{n}}S^y_{\mathbf{n+a}}
+\lambda S^z_{\mathbf{n}} S^z_{\mathbf{n+a}} \Bigr] .
\end{equation}
The XY-model results for $\lambda=0$.
Values of $\lambda$ below $1$ describe a system where $z$ is the hard axis 
and $xy$ is the easy plane, allowing for the appearance of vortices.
The total number of spins in the system is 
\begin{equation}
N=N_{\textrm{occ}} = (1-\rho_{\textrm{vac}}) L^2.
\end{equation}
In general, calculated thermodynamic quantities are quoted here 
as per-occupied-site average values, i.e., normalized by $N$.


\section{MC Simulations} 
\label{MC}
Classical Monte Carlo algorithms were used to estimate static
thermodynamic quantities as functions of temperature $T$, 
with emphasis on the internal energy 
$e=E/N$, specific heat $c=C/N$, and magnetic susceptibility of the 
in-plane spin components, $\chi$, all per-occupied-site quantities,
as well as the vorticity densities per occupied site.
It is understood that a certain percent of vacancies $\rho_{\textrm{vac}}$
has been produced in the $L\times L$ lattice under study, at randomly
selected positions as described above.
We found there to be very little variation in the results with the choice of
equivalent systems with different vacancy positions, especially for the
larger lattices (i.e., a large system is self-averaging).
Therefore, no averaging over different systems at a given $L$ was performed.
%

\subsection{MC Algorithm}
The MC techniques used here have been described in Ref.\ 
\onlinecite{Gouvea97} and are based partly on simulation methods
developed in Refs.\ \onlinecite{Kawabata+86a,Kawabata+86b,Wysin90,Landau99}.
We applied a combination of Metropolis single-spin moves and over-relaxation 
moves\cite{Evertz96} that modify all three spin components, and in addition, Wolff 
single-cluster operations\cite{Wolff88,Wolff89} that modify {\em only} the $xy$ 
spin components.
The single spin moves and over-relaxation moves were applied to sites
selected randomly in the lattice;  similarly, the initial sites for
cluster generation were selected randomly.

In the single-spin moves, randomly selected spins were modified by
adding small increments in random directions, and then renormalizing 
the spins to unit length, accepting or rejecting each change according 
to the Metropolis algorithm.

The over-relaxation and cluster moves are important at low temperatures, 
where the $xy$ spin components tend to freeze and single spin moves become 
inefficient.
Over-relaxation and cluster moves have the tendency to change spin 
directions with no or very small changes in energy, hence, their 
efficacy at low temperature. 

The over-relaxation moves used here consist of reflecting a randomly
selected spin across the effective magnetic field due to its neighbors,
\begin{equation}
\vec{B}_{\mathbf{n}}=J \sum_{\mathbf{a}} p_{\mathbf{n+a}}
[ S_{\mathbf{n+a}}^x \hat{x} + S_{\mathbf{n+a}}^y \hat{y} +
 \lambda S_{\mathbf{n+a}}^z \hat{z} ],
\end{equation}
while preserving the spin length.
All spin components are involved in the process, and the $z$ components
become more greatly involved when the anisotropy parameter $\lambda$ 
approaches $1$.
This spin change exactly conserves the energy, while effectively mixing
up the spin directions.

The Wolff cluster algorithm (and computer subroutine) used here is identical 
to that used for the pure system without vacancies.
In the actual computations, the spins of the vacant lattice sites are set 
to zero length (equivalent to setting occupation $p_{\mathbf{n}}=0$), 
and the calculations proceed normally.
No other significant changes are needed to implement the Wolff algorithm. 
It means that the Wolff clusters being formed could actually span across 
vacant sites.
Clearly, this means that a large cluster being formed might actually
be composed from several sub-clusters connected by vacant sites, a
situation that probably enhances the mixing produced by the algorithm.

For a single MC step (an MC pass through the lattice), we \emph{attempted} 
$N$ over-relaxation moves, followed by $N$ single-spin moves, 
followed by $N$ cluster moves.
An initial set of 5000 MC steps was used to equilibrate the system.
The averages shown here result from a sequence of 300,000 MC steps at
each individual lattice size and temperature.
For most of the data, the error bars are smaller than the symbols used,
hence, error bars have not been displayed. 

\subsection{MC Measurements}
In terms of temperature $T$ and Boltzman's constant $k$, the  system's
thermodynamic energy $E$  and heat capacity $C$ are defined 
via usual relations,
\begin{equation}
E = \langle H \rangle, \quad C= k [ \langle H^2 \rangle - \langle H \rangle^2]/T^2.
\end{equation}
The instantanteous total magnetization of the system is the sum 
over all spins
\begin{equation}
\vec{M} = \sum_{\mathbf{n}} p_{\mathbf{n}} \vec{S}_{\mathbf{n}}.
\end{equation}
For purposes of finding $T_c$, it is important to calculate the
associated per-spin susceptibility $\chi^{\alpha\alpha}$ of any 
component $\alpha$, derived from the magnetization fluctuations,
\begin{equation}
\label{chiaa}
\chi^{\alpha\alpha} = (\langle M_{\alpha}^2 \rangle 
-  \langle M_{\alpha}\rangle^2)/(NT).
\end{equation}
Both $\chi^{xx}$ and $\chi^{yy}$ were computed by (\ref{chiaa}) 
and then averaged to get the in-plane susceptibility,
\begin{equation}
\chi = (\chi^{xx}+\chi^{yy})/2.
\end{equation}
Finite size scaling of $\chi$ was found to be the best method to determine
$T_c$ precisely, see below.

In the thermodynamic limit, according to the Mermim-Wagner theorem, 
$\langle \vec{M} \rangle \rightarrow 0$ at any temperature, and this
holds in an approximate sense in the MC averages of finite systems. 
Therefore it is also interesting to calculate the system's total in-plane 
absolute valued magnetic moment (order parameter $M^*$), which only
tends to zero in the high-temperature phase, 
and its associated per-spin susceptibility $\chi^*$, 
\begin{equation} 
M^* = \langle \sqrt{M_x^2+M_y^2}~ \rangle,
\quad \chi^* = [\langle M_x^2 + M_y^2 \rangle 
-  {M^{*}}^2]/(NT).
\end{equation}
Related per-spin energy, specific heat, and order parameter ($e, c, m^*$), 
are obtained by dividing each by the number of occupied sites, $N$.


\begin{figure}
\includegraphics[angle=-90.0,width=\columnwidth]{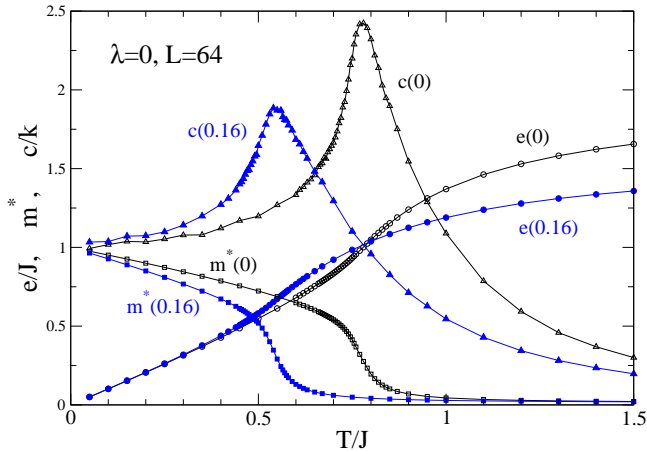}
\caption{
\label{eclam0}
For the model with edge $L=64$, at the XY limit, the internal energy,
absolute magnetization,  and specific heat per spin for the uniform system 
and with 16\% vacancy density.
}
\end{figure}

For example,  at $\lambda=0$, $L=64$, typical results for the energy,
absolute in-plane magnetization and specific heat per spin are shown in 
Fig.\ \ref{eclam0}, comparing the pure system with that at 16\% vacancy 
concentration.
Note that the energy and specific heat per spin have rather weak 
dependence on the system size $L$, while $m^{*}$ acquires a sharper
dropoff with inceasing $L$.
The most obvious effect of $\rho_{\textrm{vac}}>0$ is the lowering of
the BKT transition temperature.
A less obvious effect is the lowering of the per-spin energy and specific heat
in the high-temperature phase.
This quite possibly results because a large fraction of the vortices
produced in the high-temperature disordered phase are localized on
the vacancies, as found below. 
When thus formed, vortices require a lower nucleation energy, and the 
system can reach a specified entropy at a lower overall energy cost.
 
\subsection{Critical Temperature}
Initially, the fourth order in-plane magnetization cumulant $U_L$ due
to Binder\cite{Binder81,Binder90} was calculated to aid in location of the 
transition temperature in the thermodynamic limit.
It is defined using a ratio,
\begin{equation}
U_L = 1- \frac{ \langle (M_x^2 + M_y^2)^2 \rangle}
              {2 \langle M_x^2 + M_y^2 \rangle^2 }.
\end{equation}
This quantity becomes 0.5 in the low-temperature ordered limit,
and tends towards zero in the disordered high-temperature limit.
When measured at the critical temperature, its value is expected to
be approximately independent of the system size.  
Therefore, $T_c$ can be estimated by plotting $U_L$ vs.
$T$ for different system sizes and observing the common crossing point
of the data.  
This definition of $U_L$ is analogous to the more familiar form
that would be applied to a single in-plane spin component or
single-component model, viz.,
\begin{equation}
U_L^{(x)} = 1 - \frac{\langle M_x^4 \rangle}{3\langle M_x^2 \rangle^2 }.
\end{equation}

\begin{figure}
\includegraphics[angle=-90.0,width=\columnwidth]{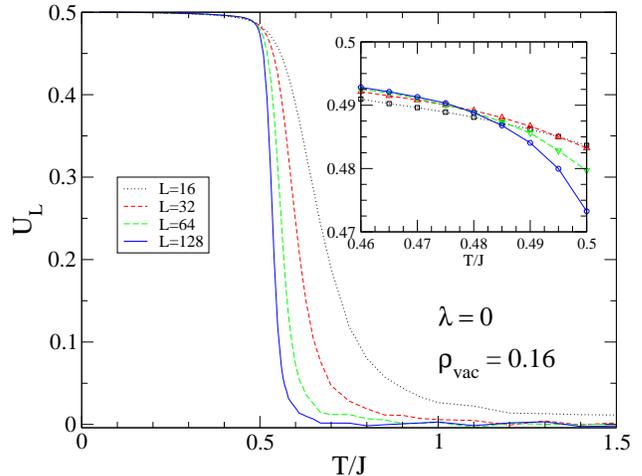}
\caption{
\label{UL16lam0}
Application of the $L$-dependence of the fourth order cumulant $U_L$
on various system sizes to estimate $T_c/J\approx 0.48$ (common crossing 
point of the data) at 16\% vacancy density in the XY model. 
}
\end{figure}

Example application of $U_L$ for finding $T_c$ for the XY model
at 16\% vacancy concentration is shown in Fig.\ \ref{UL16lam0}.
The transition temperature is lowered to $T_c \approx 0.48 J$,
considerably less than $T_c \approx 0.70 J$ that holds at
zero vacancy concentration.

It is seen, however, that $U_L$ requires an excessive amount of calculations 
even to get two-digit precision for $T_c$.
Following Cuccoli \textit{et al.}\cite{Cuccoli+95} and their analysis of the 
pure XXZ model, a finite scaling analysis of the in-plane susceptibility $\chi$ 
is seen to be much more precise and efficient for finding $T_c$.
The essential feature needed here is that near and below $T_c$, the susceptibility 
scales with a power of the system size,
\begin{equation}
\chi \propto L^{2-\eta},
\end{equation}
where the exponent $\eta$ describes the long distance behavior of in-plane spin 
correlations below $T_c$, see Ref.\ \onlinecite{Cuccoli+95} for details.
Importantly, at the transition temperature for the XY model, one has $\eta=1/4$.
Here we make the assumption that $\eta=1/4$ at $T_c$ also for the models with $\lambda>0$
and with vacancies present.
The validity of this assumption is partially tested by the quality of the scaling 
that it produces.

\begin{figure}
\includegraphics[angle=-90.0,width=\columnwidth]{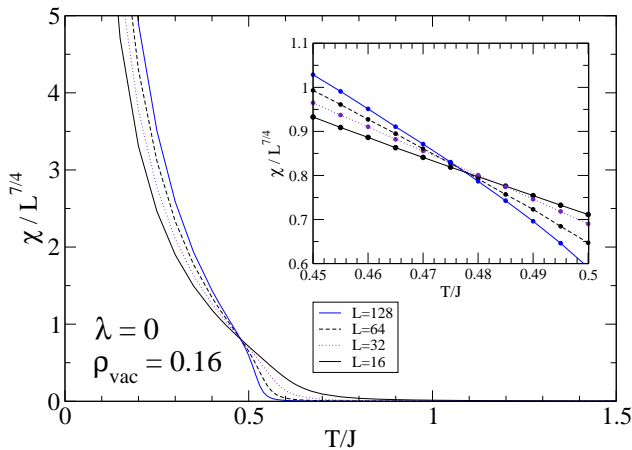}
\caption{
\label{chi_lam0_16}
Application of the finite-size scaling of in-plane susceptibility 
to estimate $T_c/J \approx 0.478$ (common crossing point of the data) 
at 16\% vacancy density in the XY model, using exponent $\eta=1/4$. 
}
\end{figure}

Using $\eta=1/4$, we plotted $\chi/L^{7/4}$ versus $T$ for the data from
different system sizes, $L=16, 32, 64, 128$ together on one graph.
The common crossing point of the curves locates the critical temperature,
for example, the result $T_c/J \approx 0.699 \pm 0.001$ is easily reproduced
for the vacancy-free XY model.
An example of this is given in Fig.\ \ref{chi_lam0_16}, for $\lambda=0$
at 16\% vacancy concentration.
An exceptionally tight crossing point occurs at the critical temperature,
$T_c/J \approx 0.478  \pm 0.001 $.  
The clarity of the crossing point gives considerable confidence in the 
$\eta=1/4$ assumption, even when vacancies are present.  
Similar results hold for the other models studied (nonzero $\lambda$
and nonzero $\rho_{\textrm{vac}}$, see \ref{others}) where the scaling 
estimates of $T_c$ give dramatic improvement upon the more approximate 
estimates using $U_L$, from the \emph{same} MC data.

\subsection{Vortex densities}
In a system with vacancies, the presence of unit charged and doubly charged 
vortices is determined as follows.

If a unit cell or plaquette is found to be fully occupied by spins, then the 
vortex search takes place in the usual way, counting the net vorticity there
by summing the in-plane angular changes around the cell and normalizing
by $2\pi$: 
\begin{equation}
\label{qdef}
q=\frac{1}{2\pi} \sum_{\textrm{edge bonds}} \Delta\phi_{\textrm{bond}}.
\end{equation}
It is understood that each difference between two in-plane spin angles 
along one edge segment must be taken on the primary branch:
$-\pi/2 < \Delta \phi_{\textrm{bond}} < \pi/2$.  
Then $q$ within a cell is forced to be an integer.  
In practice, the possible outcomes for $q$ are 0 and $\pm1$, as higher charged
vortices are unstable within a single cell of the lattice, and never
occur in Monte Carlo simulations.

Additionally, the search for vorticity can also be performed easily
around the quartet of unit cells that surrounds an individual vacancy.
A vacancy is surrounded automatically by eight occupied sites,
connected by eight bonds (under our assumption of repulsive vacancies).  
Then again Eq.\ (\ref{qdef}) can be applied to determine the total
vorticity within these four cells nearest the vacancy, summing over 
the in-plane angular changes in all eight bonds.
Now it is seen that the result for $q$ can take the additional 
possible values $q=\pm2$, i.e., \textit{doubly charged vortices}
are found to be stable entities when localized on the vacancies,
but never are found to occur separated from a vacancy.

\begin{figure}
\includegraphics[width=\smallfigwidth]{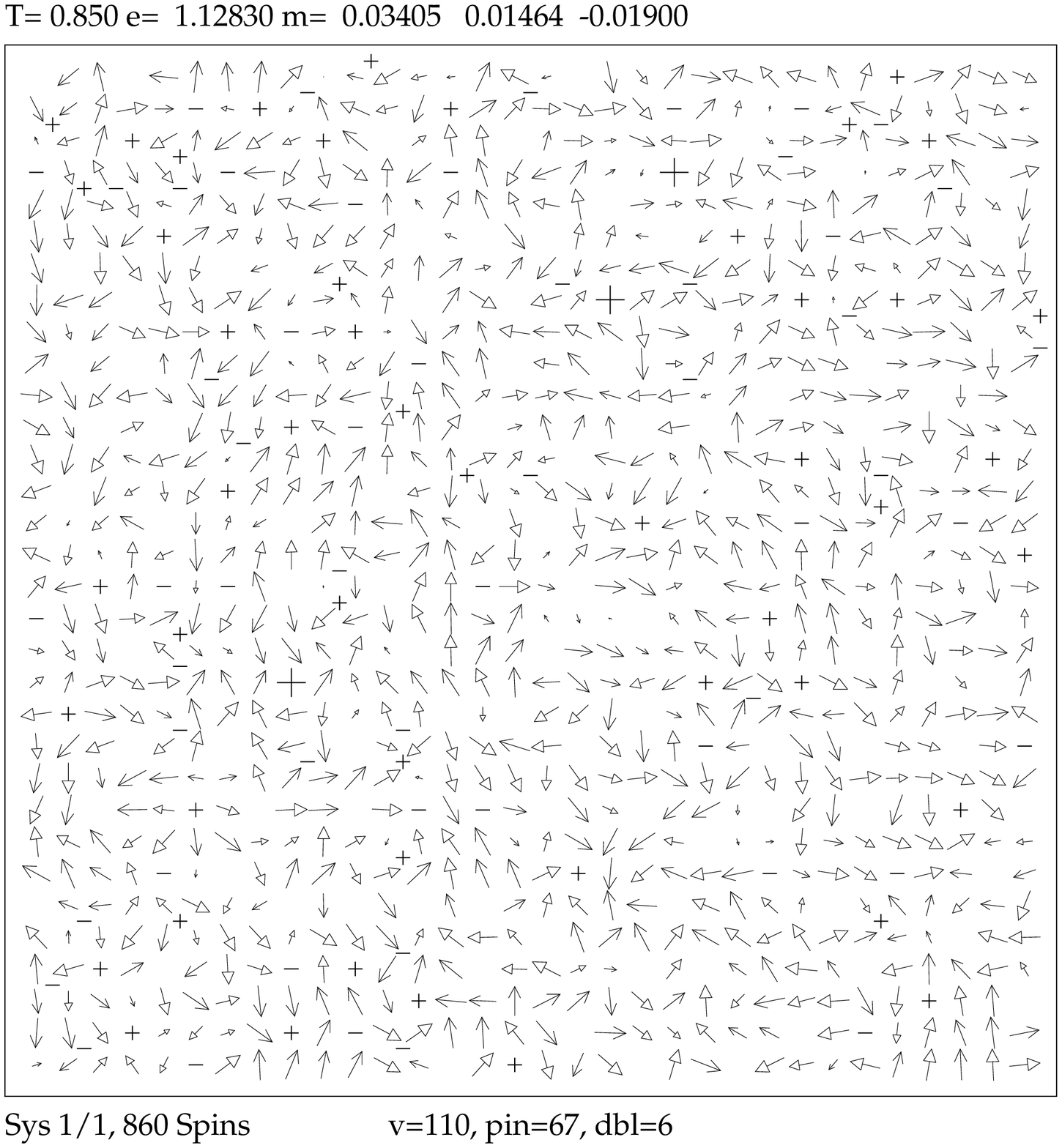}
\caption{
\label{spin32x32T85}
A spin configuration from MC simulations for $L=32$, $\lambda=0$, 
$\rho_{\textrm{vac}}=0.16$, at $T=0.85 J$, with vortices indicated by $\pm$ signs.  
The projections of the $xy$ spin components are shown as arrows, with line and 
triangular heads indicating positive/negative $z$ spin components.  The three 
larger plus signs are vortices of charge $q=+2$ centered on vacancies.  Many 
vortices have formed centered on vacancies.  
}
\end{figure}

An example of a state with doubly charged vortices is given in Fig.\
\ref{spin32x32T85}, produced in the MC simulations with $L=32$,
$\lambda=0$, $\rho_{\textrm{vac}}=0.16$, at $T=0.85 J$, well above 
the critical temperature ($T_c\approx 0.478 J$) for this vacancy
concentration.
The locations of the $q=2$ vortices are indicated by the larger 
plus signs; two near the top-center and one in the lower-left section
of the system.  
Other singly charged vortices are indicated by the smaller $\pm$ signs.
One can also note the considerable number of vortices (of any charge)
that form exactly centered on the vacancies. 

For a state in which there are $n_1$ singly charged vortices ($q$ either 
+1 or -1) and $n_2$ doubly charged vortices ($q$ either +2 or -2), the
total absolute vorticity density was defined relative to the occupied 
spin sites, and giving a double weight to the double charges,
\begin{equation}
\rho = \frac{\sum_i \vert q_i \vert}{N} = \frac{n_1+2n_2}{N}.
\end{equation}
Additionally, the vorticity fraction $f_{\textrm{dbl}}$ that corresponds to
doubly-charged vortices was tracked,
\begin{equation}
f_{\textrm{dbl}} = \frac{2n_2}{n_1+2n_2}.
\end{equation}
%

Indeed, both the $q=\pm1$ and $q=\pm2$ vortices are commonly found
centered on the vacancies.  
Therefore, we also calculated the fraction $f_{\textrm{pin}}$ of the 
total absolute vorticity that is found centered on vacancies, or, pinned
on the vacancies:
\begin{equation}
f_{\textrm{pin}} = \frac{\sum_i \vert q_i^{(\textrm{pinned})} \vert }
                        {\sum_i \vert q_i \vert },
\end{equation}
where the sum in the denominator is over all vortices found in the
system.
As already mentioned above, the doubly charged vortices are always found
pinned on the vacancies.
Furthermore, at low temperatures with very low vortex density, essentially
all vortices nucleate on vacancies.

\begin{figure}
\includegraphics[angle=-90.0,width=\columnwidth]{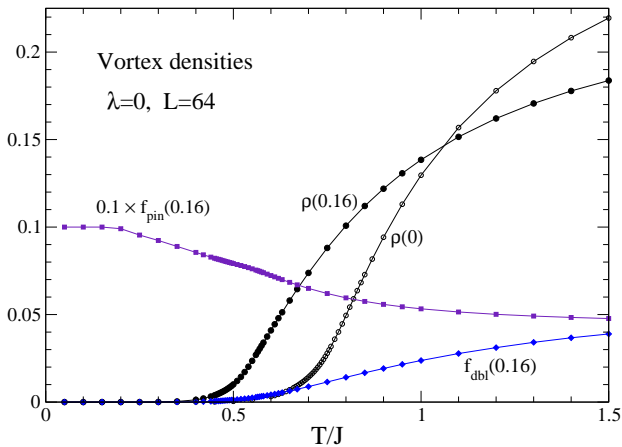}
\caption{
\label{rho_lam0}
Thermally induced vorticity density for the uniform XY model [$\rho(0)$]
and at 16\% vacancy density [$\rho(0.16)$].  Also displayed are the
vorticity fraction pinned on vacancies [$f_{\textrm{pin}}$]
and the fraction with doubled charges [$f_{\textrm dbl}$], both
when $\rho_{\textrm{vac}}=0.16$.
}
\end{figure}

Typical results for these various vorticity densities in the XY model at $L=64$
are shown in Fig.\ \ref{rho_lam0}.
Considering the curves for 16\% vacancy concentration, it is significant that 
for temperatures near $T_c$, the pinned vorticity fraction is
around 75\%.
This is reasonable, because pinned $q=1$ vortices have considerably lower
energy than free ones and therefore will dominate at the lower temperatures.
On the other hand, doubly-charged vorticity does not appear with significant 
population until well into the high-temperature phase, when it composes
up to several percent of the total vorticity in the system.

\subsection{Variations with $\lambda$}
\label{others}
The previous sections presented vacancy effects in the XY model, $\lambda=0$.
MC simulations were also carried out at two nonzero values of the
anisotropy parameter:  1) the vortex-in-plaquette critical anisotropy 
($\lambda_{\textrm{c}}=0.70$) and 2) the vortex-on-vacancy
critical anisotropy ($\lambda_{\textrm{cv}}=0.9545$).
At $\lambda_{\textrm{c}}$ large out-of-plane magnetization fluctuations 
might be expected if free vortices were dominating the dynamics.
At $\lambda_{\textrm{cv}}$ large out-of-plane magnetization 
fluctuations might be expected if vortices pinned on vacancies
were dominating the dynamics.

At these nonzero $\lambda$, the effects due to vacancies are similar to 
those found at $\lambda=0$: reduction of $T_c$, significant 
fraction of pinned vorticity in the low-temperature phase, and appearence 
of doubly charged vorticity in the high-temperature phase.

These limited results for $T_c$ as determined by scaling of $\chi$ are 
summarized in Table \ref{TcTable}.
At 16\% vacancy concentration, the general dependence of $T_c$ on $\lambda$
mimics that found for the pure model;  $T_c$ changes very little until
$\lambda$ becomes very close to $1$.

\begin{table}
\caption{\label{TcTable} Dependence of critical temperature $T_c(\rho_{\textrm{vac}})$
 on anisotropy constant $\lambda$, for the pure model ($\rho_{\textrm{vac}}=0$) 
and at $\rho_{\textrm{vac}}=0.16$, obtained by the scaling of in-plane
susceptibility.}
\begin{ruledtabular}
\begin{tabular}{cccc}
$\lambda$  &  $T_c(0)/J$  & $T_c(0.16)/J$ \\
\hline
0.0  &  $0.699  \pm 0.001$  &   $0.478 \pm 0.001$ \\
0.7  &  $0.673  \pm 0.001$  &   $0.454 \pm 0.001$ \\
0.9545 & $0.608 \pm 0.001$  &   $0.404 \pm 0.001$ \\
\end{tabular}
\end{ruledtabular}
\end{table}

\begin{figure}
\includegraphics[angle=-90.0,width=\columnwidth]{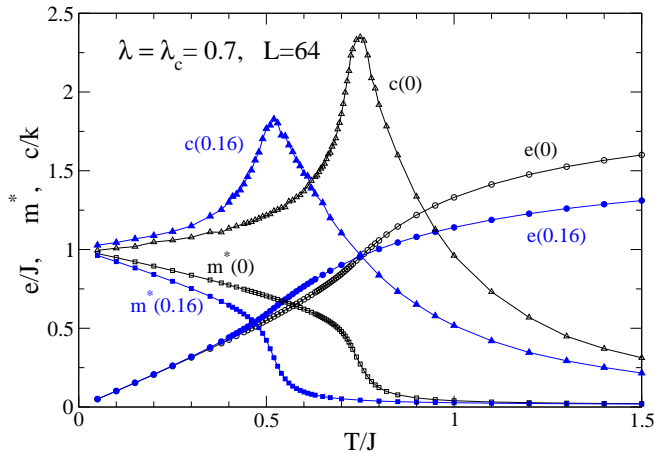}
\caption{
\label{eclam7}
For the model with edge $L=64$, at the vortex-in-plaquette critical
anisotropy, the internal energy, absolute magnetization, and specific heat 
per spin for the uniform system and with 16\% vacancy density.
}
\end{figure}
The per-spin energy, absolute in-plane magnetization, and specific heat 
at $\lambda=\lambda_c$ are shown in Fig.\ \ref{eclam7}, where a mildy 
different result is seen compared to the XY model.
\begin{figure}
\includegraphics[angle=-90.0,width=\columnwidth]{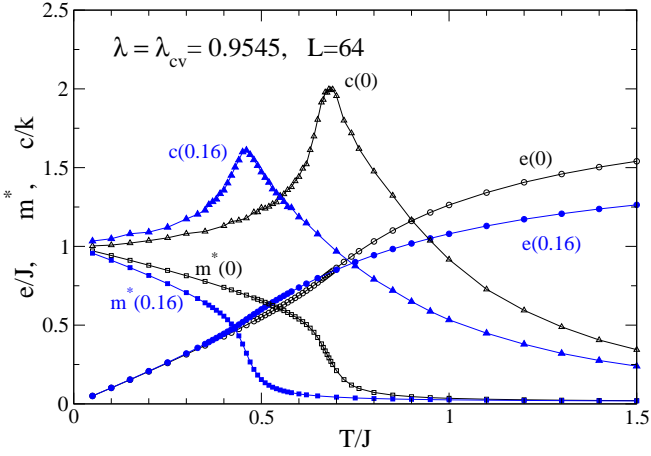}
\caption{
\label{eclamcv}
For the model with edge $L=64$, at the vortex-on-vacancy critical anisotropy, 
the internal energy, absolute magnetization and specific heat per spin for 
the uniform system and with 16\% vacancy density.
}
\end{figure}

\begin{figure}
\includegraphics[angle=-90.0,width=\columnwidth]{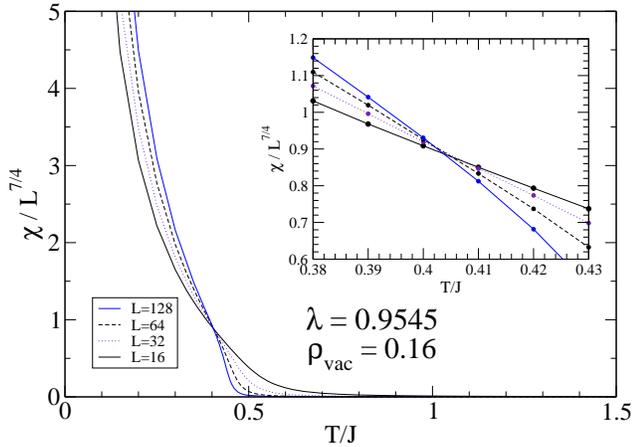}
\caption{
\label{chi_lamcv_16}
Application of the finite-size scaling of in-plane susceptibility 
to estimate $T_c/J \approx 0.404$ (common crossing point of the data) 
at 16\% vacancy density at the vortex-on-vacancy critical anistropy.
}
\end{figure}

\begin{figure}
\includegraphics[angle=-90.0,width=\columnwidth]{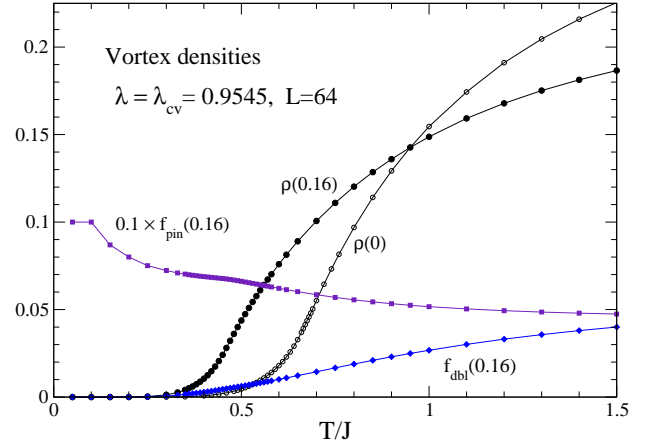}
\caption{
\label{rho_lamcv}
Thermally induced vorticity density [$\rho(0)$] at the vortex-on-vacancy 
critical anisotropy with 16\% vacancy density [$\rho(0.16)$].  Also 
displayed are the vorticity fraction pinned on vacancies 
[$f_{\textrm{pin}}$] and the fraction due to double charges 
[$f_{\textrm dbl}$], at 16\% vacancy density.
}
\end{figure}

\begin{figure}
\includegraphics[angle=-90.0,width=\columnwidth]{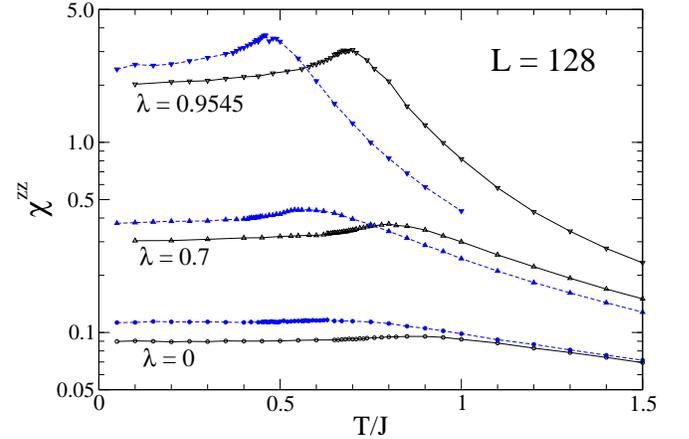}
\caption{
\label{chiz}
Out-of-plane susceptibilities $\chi^{zz}$ vs. temperature for
$L=128$, at the three anisotropies studied.  The lower curves at
low $T$ (open symbols) correspond to $\rho_{\textrm{vac}}=0$, the 
upper curves (solid symbols) correspond to $\rho_{\textrm{vac}}=0.16$ .
Unlike $\chi^{xx}$ or $\chi^{yy}$, there is only a very weak dependence 
of $\chi^{zz}$ on $L$, mostly in the high-temperature phase.
}
\end{figure}

At $\lambda_{\textrm{cv}}$, stronger effects are found, as seen in 
Fig.\ \ref{eclamcv}.
The transition temperature is reduced to $T_c/J \approx 0.404$
when 16\% vacancies are present, Fig.\ \ref{chi_lamcv_16}, compared to 
$T_c/J \approx 0.608$ for the pure system.
The vorticity density results are shown in Fig.\ \ref{rho_lamcv},
and mimic those found for the XY model.
Comparing the results at the different anisotropies, there is no
sudden change in the vacancy effects, as far as can be seen from 
these limited data.
The out-of-plane fluctuations vs. $T$ for these nonzero $\lambda$
do not exhibit any particularly significant features due to the 
presence of vacancies. 
Generally, in the low-temperature phase, $\chi^{zz}$ increases with 
vacancy density, but even more so with increasing $\lambda$, as summarized 
in Fig.\ \ref{chiz}.
It is clear that the out-of-plane fluctuations are aided by the presence 
of vacancies, but from the limited data here, no significant conclusion
about the role of pinned vortices vs. free vortices can be drawn.
%

\section{Doubly-charged vortex configurations from spin relaxations}
Having seen the appearence, in general, of doubly charged vorticity 
localized on the vacancies, it is important to consider the basic
analysis of their energetics. 
Clearly, in continuum theory, the static vortex energy (dependent on an 
integral of the form 
$ J \int d^2x ~ |\nabla \phi|^2 \approx J \pi q^2 \ln(R/a)$ ) 
is proportional to the squared charge.
Therefore, one expects that the doubly-charged vortices, even when pinned
on vacancies, should have considerably higher energy than singly-charged
vortices (either pinned or free).
Apparently, the absence of a spin at the center of the $q=2$ vortex,
and the missing four interior bonds, significantly reduces the energy and 
allows for stability.

Following the procedure in Ref.\ \onlinecite{Wysin03}, some doubly-charged 
vortex configurations were investigated numerically for their stability
as a function of the anisotropy.
For simplicity, a circular system of radius $R$, with sites defined
on a square lattice, was used.
The vacant site was placed exactly at the center of the circular system.
Free boundary conditions applied at the edge of the system.
The initial in-plane spin angles were set to that for a charge $q$
vortex centered at position $(x_v, y_v)$,
\begin{equation}
\label{phiq}
\phi(x,y) = q \tan^{-1}\left(\frac{y-y_v}{x-x_v}\right) + \phi_0.
\end{equation}
For a $q=\pm 2$ vortex at the center of the system, a convenient 
way to implement this expression on the $xy$ spin components for
arbitrary constant $\phi_0=0$, without using trigonometric functions 
is
\begin{equation}
S^x(x,y) = \frac{x^2-y^2}{x^2+y^2}, 
\quad S^y(x,y)=\frac{\pm 2 xy}{x^2+y^2},
\end{equation}
the $\pm$ signs producing vortex/antivortex configurations.

In order to test the in-plane to out-of-plane stability, all out-of-plane 
spin components were given small initial values $S^z = 0.001$, thereby
biasing the spin configuration possibly to go out-of-plane along the
positive z-axis. 
After this small perturbation, all spins were normalized to unit length.
The spin configuration was relaxed iteratively by setting each spin
to point along the direction of the effective field due to its neighbors,
keeping the spin length fixed at unity.  
This leads eventually to a final configuration that is a local energy
minimum of the Hamiltonian, i.e., some form of stable configuration
evolved from the original state, in some case with vorticity still
present, and in other cases, no net vorticity.

\begin{figure}
\includegraphics[angle=-90.0,width=\columnwidth]{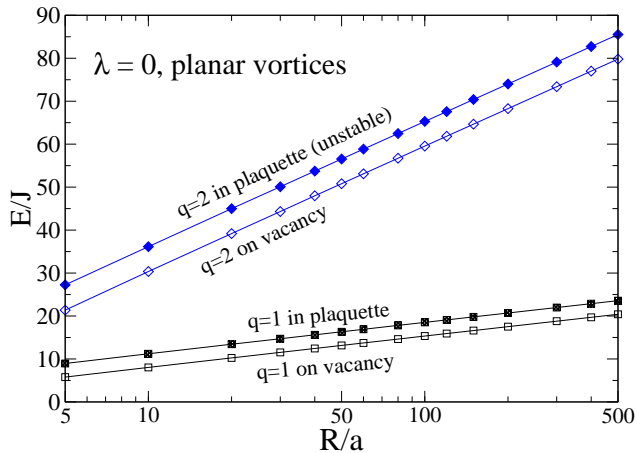}
\caption{
\label{E2R}
Various total system energies with a vortex present 
versus system radius $R$.
}
\end{figure}

\subsection{$q=2$ vortex relaxation for XY model ($\lambda=0$)}
The first numerical relaxations were applied for the XY limit,
$\lambda=0$, to get the general idea of the energy compared to
that for $q=1$ vortices.
Typically, the energy found at $\lambda=0$ should be expected to apply
rather accurately to larger values of $\lambda$, as long as the vortex
remains in the planar configuration.
These relaxations were performed for systems with radius ranging from 
$R=5a$ to $R=500a$, as the energy is expected to have a logarithmic
dependence on $R$.
The energy results $E_{\textrm{vv}}$ for a $q=2$ vortex centered on the 
vacancy are shown in Fig.\ \ref{E2R}, and compared with similar results 
for $q=1$ vortices.
Additionally, the vortex energies $E_{\textrm{vp}}$ are shown when centered 
in a plaquette. 
For $q=1$, this energy was found by relaxation to a stable vortex state, 
whereas, for $q=2$, expression (\ref{phiq}) was used to set the vortex centered 
in the plaquette, after which the energy was directly evaluated.
This latter configuration for $q=2$ is unstable, but was used for estimation
of the vortex-on-vacancy binding energy, see below.

Inspection of Fig.\ \ref{E2R} shows that, as expected, the doubly-charged
vortices have considerably higher energy compared to singly charged, and
furthermore, there is a nearly constant energy gap between the 
vortex-in-plaquette and vortex-on-vacancy states.
Each data set fits extremely well to a logarithmic dependence on $R$ in the
form $E=A+B \ln(R/a)$.
For $q=1$, both curves have slope parameter $B_1 \approx 3.17 JS^2$.
For $q=2$, both curves have slope parameter $B_2 \approx 12.7 JS^2$,
a value very close to four times as large as that for $q=1$, as might be
expected.
The extra energy requirement for the $q=2$ vortices clearly leads to a
restriction on their thermal population compared to $q=1$ vortices.
In all cases shown, the final spin configuration was found to be completely
in-plane (all $S^z=0$).

The difference between the vortex-on-vacancy and vortex-in-plaquette energies
can be taken to define an energy for binding or pinning the vortex on the vacancy,
\begin{equation}
\Delta E_q = E_{q, \textrm{vp}}-E_{q, \textrm{vv}}.
\end{equation}
Using the results shown, the binding energy for a $q=1$ vortex-on-vacancy is 
$\Delta E_1 \approx 3.177 JS^2$, using the asymptotic value as $R\to\infty$.  
For doubly charged vortices, the binding energy is moderately higher,
$\Delta E_2 \approx 5.73 JS^2$, in contrast to the considerably higher 
creation energy for $q=2$ vortices compared to $q=1$ vortices.
This result, however, must be taken with caution, since there is no
actual stable $q=2$ vortex free from a vacancy.

An alternative view of the $q=2$ vortex-on-vacancy energy would be to compare it to 
twice the energy of a system with a single $q=1$ vortex centered in a plaquette, 
($2E_{1 \textrm{vp}}$), because that is a stable state of the same total vorticity.
However, the energy of the two $q=1$ vortices, in their own isolated systems,
is always considerably less than that of a single $q=2$ vortex.  
This is because $2E_{1 \textrm{vp}}$ does not include the interaction potential
that would be present between two $q=1$ vortices within the same system,
which increases with the logarithm of their separation.
Thus it is not a good reference number for estimation of the $q=2$ binding
energy on a vacancy.

\subsection{Anisotropy dependence of $q=2$ vortex relaxation}
A preliminary analysis of the stability of a doubly-charged vortex
can be performed by looking at the dependence of the relaxed configuration
on the anisotropy parameter $\lambda \ge 0$.
It might be expected that a $q=2$ vortex could take on nonzero out-of-plane
components when $\lambda$ becomes adequately close to $1$, i.e., at weak
easy-plane anisotropy, in a manner similar to the out-of-plane crossover
for $q=1$ vortices.
The critical anisotropy could be expected to be different than the
value $\lambda_{\textrm{cv}}\approx 0.9545$ for pinned $q=1$ vortices.
Therefore, a limited number of numerical experiments were realized on
a circular system of radius $R=50a$ for various values of $\lambda$ above
zero.
Again, the initial condition was a $q=2$ vortex centered on the vacancy
at the center of the system, with small positive out-of-plane components 
($S^z=0.001$) at all sites.

\begin{figure}
\includegraphics[angle=-90.0,width=\columnwidth]{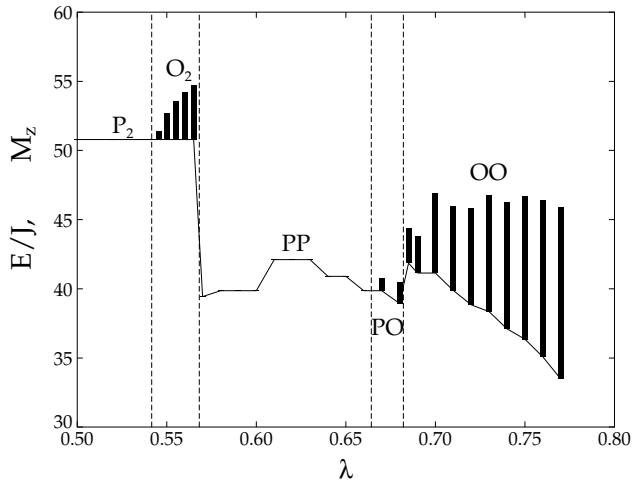}
\caption{
\label{E2vv_lambda}
After relaxation of a $q=2$ vortex initially centered on an isolated vacancy 
in a circular system of radius $R=50$, the total system energy (solid curve)
is shown as a function of the anisotropy constant $\lambda$.  The vertical
bars indicate the net out-of-plane magnetization of the relaxed configuration, 
on the same numerical scale.
}
\end{figure}

Certain aspects of these results are summarized in Fig.\ \ref{E2vv_lambda},
where the energy of the state obtained after the relaxation is plotted
versus the anisotropy parameter $\lambda$ that was used.
There are several types of results, depending on the range of $\lambda$
being considered.

For the whole range $0\le \lambda \alt 0.545$ (region $P_2$), the isolated 
$q=2$ vortex remains in a stable planar configuration on the vacancy, with no 
out-of-plane magnetization, and relatively high energy.  
For the narrow range $0.545 \alt \lambda \alt< 0.57$ (region $O_2$), the 
$q=2$ vortex remains stable on the vacancy, but develops nonzero out-of-plane 
magnetization, with an insignificant reduction in energy.
The net out-of-plane magnetization of the relaxed state is indicated in 
Fig.\ \ref{E2vv_lambda} by the bars extending above the energy curve. 
Total $M_z$ grows until $\lambda$ reaches about $0.57$, at which point the
$q=2$ vorticity concentrated on the vacancy becomes unstable, and breaks into 
one $q=1$ in-plane vortex on the vacancy, and a nearby free $q=1$ in-plane vortex.
This situation holds for $0.57 \alt \lambda \alt 0.66$ (region $PP$); the 
configuration has zero out-of-plane magnetization once again, and lower energy 
than that for the $q=2$ vortex pinned on a vacancy.
As $\lambda$ increases within this range, the free vortex progressively moves
farther from the vacancy.

\begin{figure}
\includegraphics[width=\smallfigwidth]{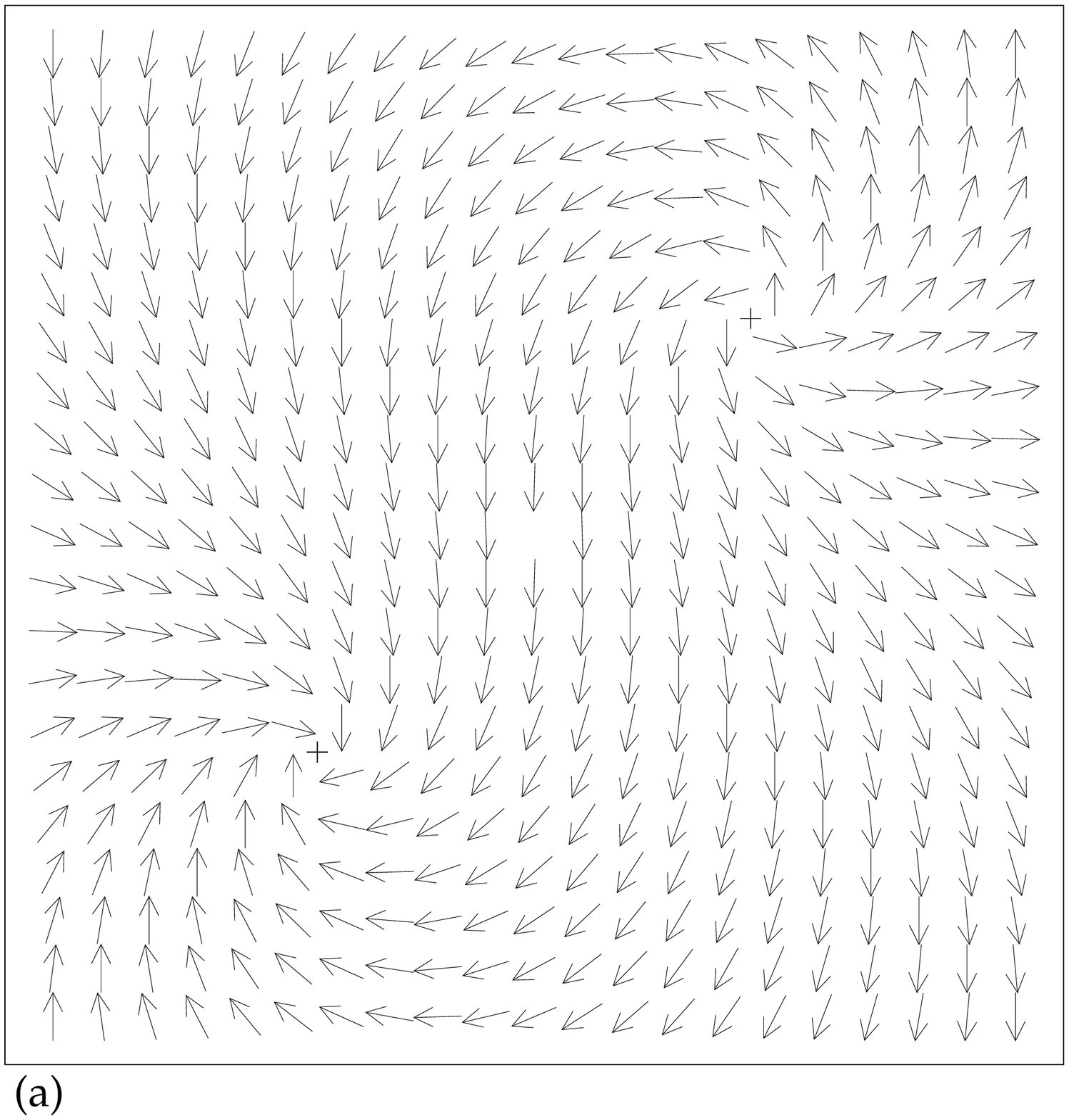}
\includegraphics[width=\smallfigwidth]{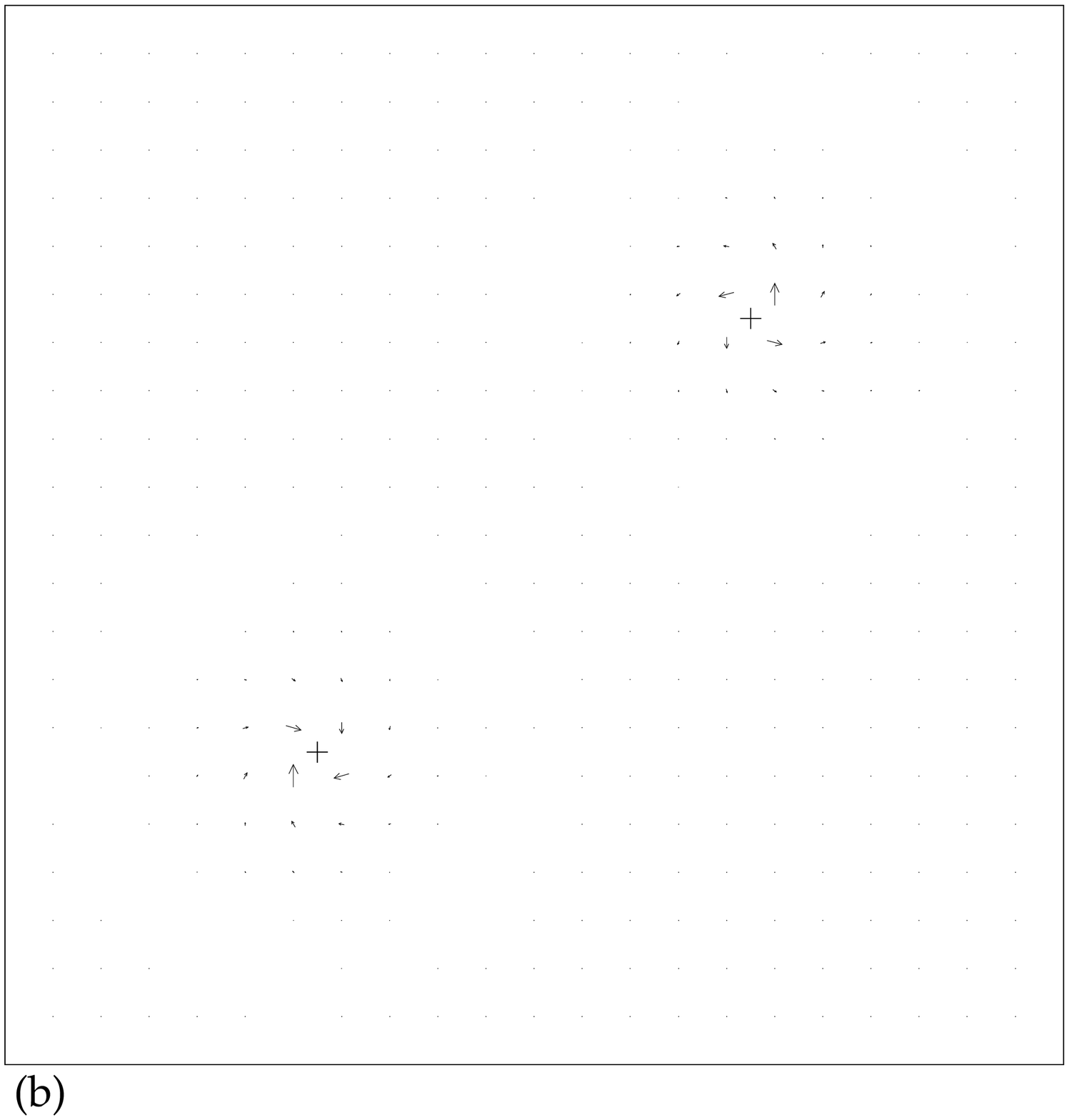}
\caption{
\label{vort7}
Final state of relaxation of a $q=2$ vortex initially centered on an 
isolated vacancy in a circular system of radius $R=50$ with $\lambda=0.7$
(only the central region of system is shown).
Part (a) shows the projection of $xy$ spin components on the plane, as
explained in Fig.\ \ref{spin32x32T85}.  In part (b) the lengths of the
arrows are equal to the $z$ spin components, while the directions are
still given by the $xy$ projections.
}
\end{figure}

When $\lambda$ ranges from about $0.67$ to $0.68$ (region $PO$), the free 
vortex starts to develop a nonzero positive out-of-plane component, while the 
pinned vortex remains planar.
Finally, at $\lambda \approx 0.685$ and above (region $OO$), the relaxed 
configuration consists of two $q=1$ positively polarized out-of-plane vortices 
centered symmetrically on opposite sides of the vacancy.  
For example, the relaxed configuration obtained for $\lambda=0.7$ is shown 
in Fig.\ \ref{vort7}.
As $\lambda$ increases, the separation of the pair increases at the same time
that their out-of-plane component increases, while the energy decreases.
Eventually, the separation surpasses the diameter of the system, and the
vorticity escapes out the boundary, leaving a final configuration with
uniform magnetization and zero energy. 
This occured for $\lambda \agt 0.77 $ in the system of radius $R=50$.

It is apparent that a localized $q=2$ vorticity has severely limited stability,
compared to $q=1$ vortices.
The $q=2$ vorticity even tends to grow out-of-plane components as a way
to enhance its stability, but this has a very limited range of utility
(region $O_2$).
Once the vorticity splits into individual $q=1$ vortices, they are seen
to influence each other, probably via an interaction with the vacancy.
This is apparent because out-of-plane components begin forming for
$\lambda$ \textit{below} the critical anisotropy parameter $\lambda_c$ 
relevant for vortices far from vacancies.
Furthermore, the pairs of out-of-plane $q=1$ vortices in region $OO$
appear to repel each other, while at the same time being attracted to
the vacancy, which would lead to a mechanically stabilized configuration.
Inspection of the spin configurations in region $OO$ (as in Fig.\ \ref{vort7})
shows spin components of one vortex to be completely symmetrical to the spin 
components in the other vortex, when reflected across the center of the system.

\section{Conclusions}
In the Monte Carlo and spin dynamics calculations presented here for
a 2D easy-plane anisotropic Heisenberg model, the presence of vacancies
has been seen to affect the details of the BKT transition and the
types of vorticity present.

\begin{figure}
\includegraphics[width=\smallfigwidth]{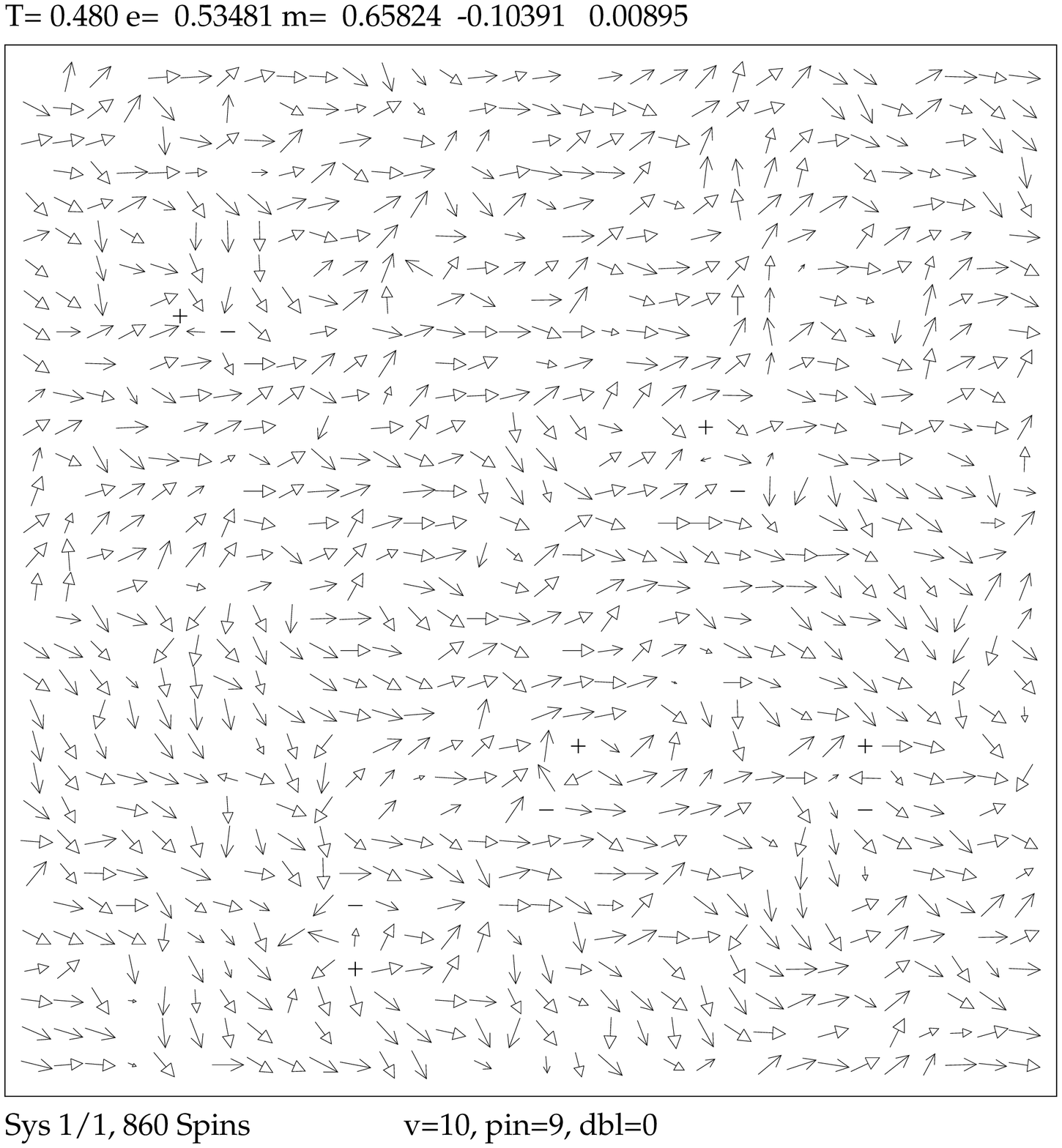}
\caption{
\label{spin32x32T48}
A spin configuration for $L=32$, $\lambda=0$, $\rho_{\textrm{vac}}=0.16$, at 
$T\approx T_c \approx 0.48 J$, where 9 out of the 10 vortices present 
have formed on vacancies.
}
\end{figure}

As seen in the planar rotator model, $T_c$ is lowered by the presence 
of vacancies.
This naturally results because the disorder in the transition becomes
dominated by the generation of vortices pinned on the vacancies.
When formed centered on vacancies, the vortex energy is significantly 
lower than that for vortices centered in plaquettes.  
Indeed, $q=1$ vortices pinned on vacancies in a square lattice, have a 
formation energy of about $3.17 JS^2$ less than when centered in plaquettes, 
while the transition temperature corresponds to an energy less than $1 JS^2$.
Thus, at temperatures near and below $T_c$, the small amount of vorticity 
that is present is predominantly pinned on vacancies, such as in Fig.\
\ref{spin32x32T48}.  
These pinned vortices are initiating and controlling the transition.
The vacancies are the nucleation sites for the spin disordering.
On the other hand, vacancies reduce the rate at which total vorticity
density rises in the high-temperature phase 
(Figs.\ \ref{rho_lam0}, \ref{rho_lamcv}).

At larger anisotropy parameter $\lambda_c$, it is known that the 
vortex-on-vacancy energy is much closer to the vortex-in-plaquette
energy.\cite{Wysin03}  
For example, at $\lambda=0.99$, 
the difference in these energies is only $0.23 JS^2$. 
Then one might expect a lesser dominance of pinned vorticity,
however, that does not appear to be the case at 
$\lambda=\lambda_{\textrm{cv}}$.
There is no substantial qualitative change in the fraction of 
pinned vorticity when compared to the XY model.
Qualitatively speaking, the details of the BKT transition at
higher $\lambda$, with vacancies, are not significantly different
than those found for the XY model.

The presence of vacancies leads to a new effect, namely, the generation
of doubly-charged vorticity that is stable when centered on vacancies.
In thermal equilibrium, this effect apparently occurs regardless of 
the easy-plane anisotropy strength.
In general, these would be thermodynamically prohibited, due to their larger 
energy, based on the usual dependence of vortex creation energy on charge 
squared.  
They still have significantly higher energy than singly charged vortices
centered on vacancies, however, the missing bonds in the core region
help to reduce their total energy compared to what they would have if
centered in a plaquette.

A spin-relaxation energy minimization shows that an individual doubly-charged 
vortex centered on a vacancy in a circular system may be stable only for a 
limited range of anisotropy constant.  
The $q=2$ vortex-on-vacancy stays stable and planar for $0\le \lambda \alt 0.545$.
In a very narrow range, $0.545 \alt \lambda \alt 0.57$, the $q=2$ vortex-on-vacancy
still remains stable, but with a small out-of-plane component.  
For $\lambda \agt 0.57$, it does not appear to be stable, but instead
breaks apart into two $q=1$ vortices that repel each other while being
attracted to the vacancy.
One might define a lower critical anisotropy 
$\lambda_{\textrm{cv,1}}\approx 0.545$ for the in-plane to out-of-plane
$q=2$ stability, and an upper critical anisotropy 
$\lambda_{\textrm{cv,2}}\approx 0.57$ for the breakdown into lower
charged vorticies.
In contrast, there is no choice of anisotropy constant that stabilizes 
a $q=2$ vortex in the center of a plaquette.

These results are intriguing, because even though they show a limited range 
of stability for doubly-charged vorticity, nevertheless, these excitations
appear in the MC simulations at $\lambda=0.7$ and $\lambda=0.9545$,
above the critical anisotropy parameters.
Of course, one could always search groups of four plaquettes in the pure model 
also to find localized vorticity of double charge (two $q=1$ vortices in
neighboring plaquettes), although it would appear very rarely, due to the 
mutual repulsion of the vortices.  
The difference here, is that the presence of a vacancy attracts vorticity
and certainly enhances the chances to find doubled vorticity within the
area of four neighboring plaquettes.  
In addition, the spin relaxations show that the doubly charged vortex can
be a static object, which can never be expected for $q=1$ vortices in
neighboring fully occupied plaquettes.

\begin{acknowledgments}
The author is very grateful for the hospitality of the Universidade 
Federal de Vi\c cosa, Brazil, where this work was completed under support 
from FAPEMIG, and for many helpful discussions there with A. R. Pereira.
\end{acknowledgments}

\bibliography{2dspins}

\end{document}